\documentclass[11pt, a4paper]{article}

\usepackage[utf8]{inputenc}
\usepackage[T1]{fontenc}

\usepackage{newtxtext, newtxmath}

\usepackage{geometry}
\usepackage{amsmath}
\usepackage{graphicx}
\usepackage{booktabs}
\usepackage{longtable}
\usepackage{array}
\usepackage{caption}
\usepackage{verbatim}
\usepackage{fvextra} 
\usepackage{pdflscape}
\usepackage[
    colorlinks=true,
    linkcolor=blue,
    urlcolor=blue,
    citecolor=blue
]{hyperref}

\title{An HTR-LLM Workflow for High-Accuracy Transcription and Analysis of Abbreviated Latin Court Hand}
\author{Joshua D. Isom \\ Lehigh University}
\date{July 5, 2025}

\begin{document}

\maketitle

\begin{abstract}
This article presents and validates an ideal, four-stage workflow for the high-accuracy transcription and analysis of challenging medieval legal documents. The process begins with a specialized Handwritten Text Recognition (HTR) model, itself created using a novel "Clean Ground Truth" curation method where a Large Language Model (LLM) refines the training data. This HTR model provides a robust baseline transcription (Stage 1). In Stage 2, this baseline is fed, along with the original document image, to an LLM for multimodal post-correction, grounding the LLM's analysis and improving accuracy. The corrected, abbreviated text is then expanded into full, scholarly Latin using a prompt-guided LLM (Stage 3). A final LLM pass performs Named-Entity Correction (NEC), regularizing proper nouns and generating plausible alternatives for ambiguous readings (Stage 4). We validate this workflow through detailed case studies, achieving Word Error Rates (WER) in the range of 2-7\% against scholarly ground truths. The results demonstrate that this hybrid, multi-stage approach effectively automates the most laborious aspects of transcription while producing a high-quality, analyzable output, representing a powerful and practical solution for the current technological landscape.
\end{abstract}


\section{Introduction}
The plea rolls of the English Court of Common Pleas (TNA series CP40) and the Court of King's Bench (KB27) represent one of the largest and most detailed sources for the social, economic, and legal history of late medieval England. As the workhorse of the English legal system, the Court of Common Pleas handled civil disputes between ordinary subjects, creating a rich panorama of daily life. However, the profound historical value of these records has been locked away by a series of formidable barriers: their sheer scale, the use of medieval Latin, and the dense, highly abbreviated "Court Hand" script, which has historically limited their accessibility to a small number of expert paleographers.

Recent advancements in AI have introduced a paradigm shift in historical document processing. While specialized HTR platforms and general-purpose multimodal Large Language Models (mLLMs) have shown impressive performance, their accuracy declines on niche, non-standard material. This article argues that for the most challenging documents, neither a single HTR model nor a direct-to-LLM approach is optimal. Instead, we propose and validate an ideal, four-stage workflow that leverages the complementary strengths of these technologies. This process uses a specialized HTR model as a stable foundation for subsequent LLM-driven refinement, expansion, and analysis, creating a practical and powerful human-in-the-loop system for unlocking these invaluable records.

\section{Related Work: LLMs in HTR}
The application of LLMs to HTR tasks is a rapidly developing field. Current approaches can be broadly categorized into three groups: LLMs as end-to-end transcription engines, as post-correction assistants, and for Named Entity Recognition (NER).

The potential of mLLMs as end-to-end transcribers has been powerfully demonstrated by tools like "Transcription Pearl" (Humphries et al. \cite{Humphries2024}), which achieved a Character Error Rate (CER) of 5.7-7\% on 18th/19th-century English documents. This performance surpassed traditional HTR tools, a finding echoed by Kim et al. \cite{Kim2025} for 20th-century French records and by Greif et al. \cite{Greif2025} for historical German printed texts.

An even more powerful application lies in multimodal post-correction, a technique pioneered for historical texts by Greif et al. \cite{Greif2025}. By providing the LLM with both the noisy OCR text and the original source image, they achieved a drastic improvement in accuracy to <1\% CER. This gold-standard performance was independently confirmed by Humphries et al. \cite{Humphries2024}, who used LLMs to correct Transkribus outputs to a word-level accuracy of 96.5\%.

Despite these successes, broad benchmarking by Crosilla et al. \cite{Crosilla2025} provides crucial context, revealing that LLM performance is highly variable across different languages and historical periods. This work, along with others (Humphries et al., \cite{Humphries2024}), also confirms that LLMs are generally incapable of effective self-correction.

Our project synthesizes these findings. We leverage the demonstrated power of LLMs for high-quality transcription (Humphries et al., \cite{Humphries2024}; Kim et al., \cite{Kim2025}) while acknowledging their biases and limitations (Crosilla et al., \cite{Crosilla2025}). Our methodology presents a fourth path: LLM-powered ground truth curation. Instead of using the LLM as the final tool, we use it to bootstrap a more stable, specialized, and open-source-based HTR model.

\section{Methodology: A Four-Stage Transcription and Analysis Workflow}
\begin{figure}[!ht]
    \centering
    \includegraphics[width=0.68\textwidth]{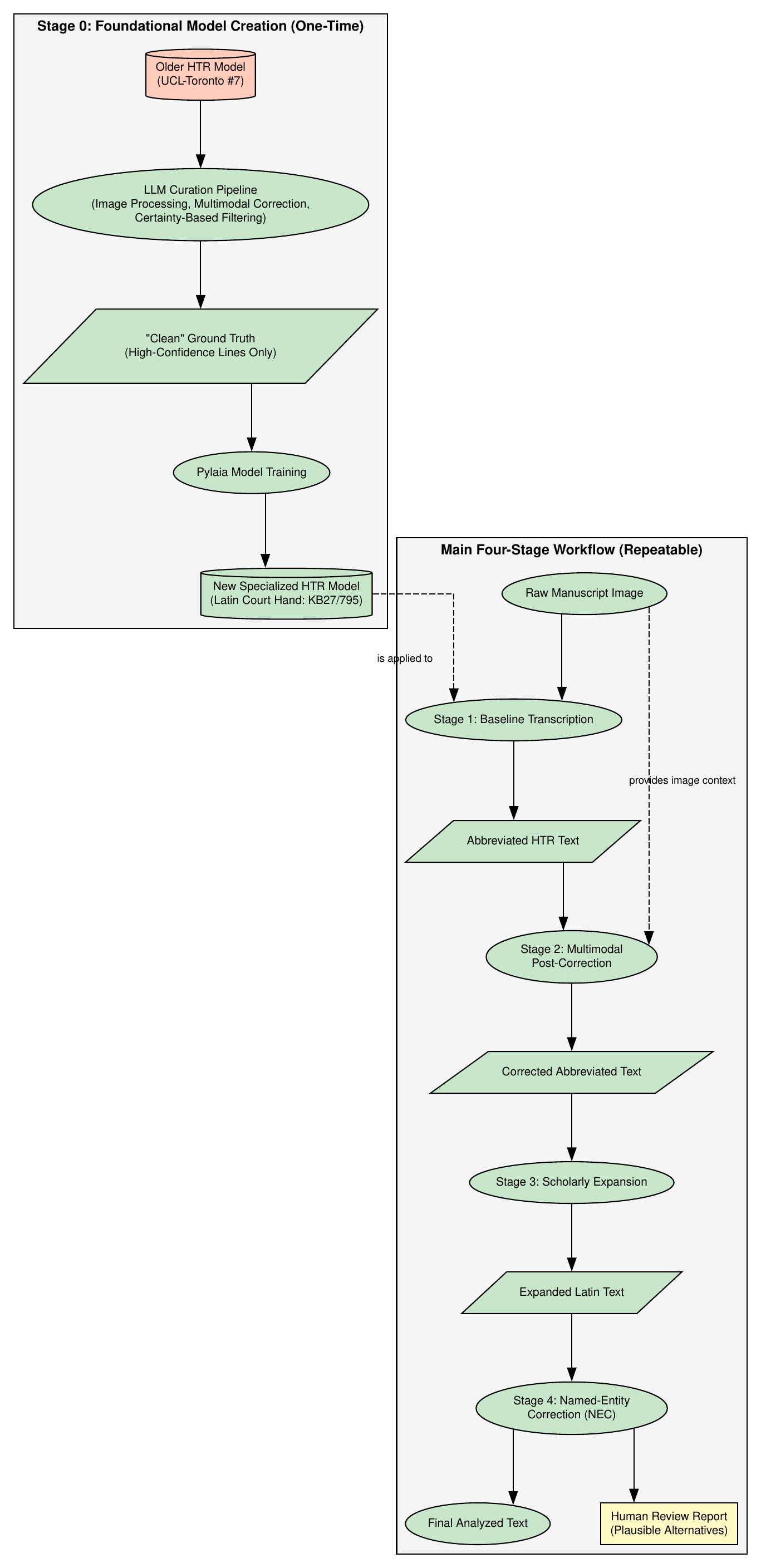}
    \caption{The Four-Stage HTR-LLM Workflow. The process begins with a one-time creation of a specialized HTR model (Stage 0), which then serves as the foundation for the repeatable four-stage pipeline that processes new documents.}
    \label{fig:workflow}
\end{figure}

Our methodology is structured as a complete research pipeline, moving from raw image to analyzable text. It begins with the creation of a foundational tool and then proceeds through four distinct stages of processing, as illustrated in Figure \ref{fig:workflow}.

\subsection{Stage 0: Building the Foundation - The "Clean GT" HTR Model}
The prerequisite for our workflow is a high-quality baseline HTR model specialized for the target script. Creating such a model traditionally requires thousands of lines of costly and time-consuming manual transcription. To circumvent this bottleneck, we developed an iterative, LLM-driven curation process to bootstrap a new model from an older, less accurate one.

The process began with a pre-existing, general-purpose HTR model for 15th-16th century English legal documents, "UCL–University of Toronto \#7,"\footnote{The model, "UCL–University of Toronto \#7," was published by the Bentham Project (University College London) and the DEEDS-project (University of Toronto) on December 13, 2022. It is publicly available in Transkribus as model ID 48734 at: \url{https://app.transkribus.org/models/public/text/48734}.} which served as our initial baseline. While functional, this model produced a high number of errors on our specific target documents. Our goal was to use this noisy output to generate a pristine ground truth (GT) set for training a new, specialized model. This was achieved through an automated pipeline executed by a Python script:

\begin{enumerate}
    \item \textbf{Initial HTR Application:} The older model was first run on our target manuscript roll (KB27/795) to generate a baseline, error-prone transcription in PAGE XML format.

    \item \textbf{Automated Image Pre-processing:} To optimize the input for the LLM, the script processed the document image in small chunks (typically 10 lines). For each chunk, it performed several key operations:
    \begin{itemize}
        \item \textbf{Cropping:} It calculated a precise bounding box around the lines in the chunk and cropped that section from the high-resolution page image, isolating the relevant text.
        \item \textbf{Annotation:} It overlaid crucial visual aids directly onto the cropped image. High-contrast red baselines were drawn for each line, and the original line number was rendered at the start and end of its corresponding baseline. This explicitly links the visual line to its numerical identifier, enforcing strict line integrity.
        \item \textbf{Deskewing:} It calculated the average angle of the baselines within the chunk and rotated the image to align the text horizontally, simplifying the visual recognition task for the LLM.
    \end{itemize}

    \item \textbf{Multimodal LLM Post-Correction:} The script then submitted each pre-processed image chunk to Anthropic's Claude 3.7 Sonnet. The multimodal prompt included the annotated image, the noisy HTR text from the older model, and a detailed system prompt (see Appendix~\ref{app:postcorrect}) instructing the LLM to produce a diplomatically faithful, non-expanded transcription. The inclusion of the old HTR text served to "ground" the LLM, constraining its task from open-ended transcription to guided correction.

    \item \textbf{Certainty-Based Curation (The "Clean GT" Innovation):} Acknowledging that LLMs are poor at self-correction, we used output variance as a proxy for uncertainty. For every image chunk, the script made two identical, independent calls to the LLM (Run A and Run B). A line was programmatically flagged as "unclear" if it met either of two conditions:
    \begin{itemize}
        \item The Character Error Rate (CER) between Run A and Run B was high (e.g., > 5\%). This indicates the model was not confident and produced different outputs for the same input.
        \item The CER between both LLM runs and the original HTR text was high. This flags lines where the LLM made substantial corrections, which have a higher probability of being speculative, especially on damaged or illegible text.
    \end{itemize}
    Lines that did not meet these conditions were considered "clear" or high-confidence transcriptions.

    \item \textbf{New Model Training:} Finally, the LLM-generated transcriptions were used as the ground truth to train a new, specialized Pylaia HTR model in Transkribus. Crucially, during the training process, all lines that had been flagged as "unclear" were ignored. This focused the model's learning exclusively on the high-confidence "Clean GT," preventing its statistical weights from being polluted by noise or speculative readings.
\end{enumerate}

This iterative process resulted in the creation of our foundational HTR model, "Latin Court Hand: KB27/795 (1460)," which achieved a CER of 5.25\% on its validation set.\footnote{The final model, "Latin Court Hand: KB27/795 (1460)," was published on May 9, 2025, and is publicly available in Transkribus as model ID 336333 at: \url{https://app.transkribus.org/models/public/text/336333}.} This stable and specialized model forms the starting point for the main four-stage workflow detailed in this paper.

\subsection{Stage 1: Baseline Transcription with a Specialized HTR Model}
The first step of the active workflow is to process the raw document images with our specialized HTR model. This provides a fast, stable, and consistent baseline transcription. The output is a diplomatic, non-expanded text that captures the abbreviations as seen on the page. While highly accurate for a single model, this HTR output contains residual errors and serves as the input for the next stage.

\subsection{Stage 2: Multimodal Post-Correction with a Grounded LLM}
This stage aims to correct the HTR baseline. We provide the LLM with both the \textbf{document image chunk} and the corresponding \textbf{HTR text} from Stage 1, using a detailed prompt to guide the process (see Appendix \ref{app:postcorrect}). This multimodal approach is crucial:
\begin{itemize}
    \item The \textbf{image} remains the absolute source of truth for the LLM.
    \item The \textbf{HTR text} acts as a "grounding" mechanism. It constrains the LLM's task from open-ended transcription to guided correction, significantly reducing the risk of hallucination and focusing its powerful analytical capabilities on fixing specific errors within a given structure. This leverages the LLM's broad contextual knowledge, which far exceeds that of the HTR model, to resolve ambiguities the HTR model could not. The output is a corrected, but still abbreviated, transcription.
\end{itemize}

\subsection{Stage 3: Scholarly Expansion of Abbreviations}
With a clean, abbreviated text, the next step is expansion. The LLM is given the corrected text from Stage 2 and a detailed prompt (see Appendix \ref{app:expansion}) that specifies the exact scholarly conventions for expanding Latin abbreviations (e.g., \texttt{p'd'co} to \texttt{predicto}, \texttt{\&c} to \texttt{etc.}). This stage transforms the diplomatic transcription into a fully readable, unabbreviated Latin text suitable for analysis.

\subsection{Stage 4: Named-Entity Correction and Alternative Generation}
The final stage focuses on the most critical data points for historical research: proper nouns. The expanded text from Stage 3 is passed to the LLM with a specialized Named-Entity Correction (NEC) prompt (see Appendix \ref{app:nec}). This prompt instructs the LLM to act as a paleographic and onomastic expert, specifically tasked with:
\begin{enumerate}
    \item Identifying all persons and places.
    \item Correcting systematic HTR errors (e.g., mistaking a capital 'B' for a 'G').
    \item Generating a list of plausible alternative readings for ambiguous names, complete with a confidence score.
\end{enumerate}
The final output is a clean, expanded text where entities have been regularized, and a "human-in-the-loop" report that flags potential ambiguities for expert review.

\section{Results: Validating the Four-Stage Workflow}
To validate the performance of this four-stage workflow, we conducted detailed analyses on four validation test cases against published scholarly ground truths, primarily drawn from J. H. Baker's edition of \textit{The Reports of John Caryll} \cite{Baker1999}. The results from four representative cases are presented below. Two cases are presented in detail to illustrate the workflow's output, while two are summarized.

\subsection{Case Study 1: Analysis of a Maintenance Plea (Summary)}
The first case, a 15th-century plea of maintenance from CP40/922~\cite{waalt-cp40-922}, achieved a final WER of approximately 6.3\%. While the workflow successfully identified most entities, the analysis revealed critical failure modes. These included semantic errors in common nouns (altering the subject of the dispute from \texttt{detentionis unius boviculi} to \texttt{detencionis bonorum}), the complete failure to reconstruct a standard legal formula (\texttt{Et defendit vim et injuriam...}) from heavily corrupted HTR, and the inability to generate the correct name (\texttt{Calney}) as a plausible alternative for the HTR's reading.


\subsection{Case Study 2: Analysis of a Trespass Plea}
To test the workflow's robustness against different scribal hands, a second validation case was analyzed: a 15th-century plea of trespass from CP40/906~\cite{waalt-cp40-906}. The final WER was approximately 2.0\% when compared against the published scholarly ground truth~\cite{Baker1999}. The NEC step was pivotal, correctly identifying a systematic HTR error (\texttt{Grome} to \texttt{Brome}). However, the analysis also revealed critical limitations. Table~\ref{tab:case2} provides a complete, unabridged alignment of the raw HTR output and the three subsequent stages of LLM processing.

\begin{landscape}
\small
\centering
\begin{longtable}{>{\raggedright\arraybackslash}p{0.23\linewidth} >{\raggedright\arraybackslash}p{0.23\linewidth} >{\raggedright\arraybackslash}p{0.23\linewidth} >{\raggedright\arraybackslash}p{0.23\linewidth}}

\caption{Aligned Transcription Comparison for a 15th-Century Trespass Case. Key: \textbf{Bold} indicates a significant error, correction, or alteration. \textit{Italics} indicate a minor variation or an acceptable medieval spelling. Plain text matches ground truth.}  
\label{tab:case2} \\

\toprule
\textbf{Raw HTR} & \textbf{LLM-Corrected Abbr.} & \textbf{Initial LLM Expansion} & \textbf{Final LLM Expansion (NE Corrected)} \\
\midrule
\endfirsthead

\caption[]{-- \textit{Continued from previous page}} \\
\toprule
\textbf{Raw HTR} & \textbf{LLM-Corrected Abbr.} & \textbf{Initial LLM Expansion} & \textbf{Final LLM Expansion (NE Corrected)} \\
\midrule
\endhead

\bottomrule
\endfoot

\bottomrule
\endlastfoot

Will'mus Hyotte nup' de \textbf{Tewdewesbury} in com' p'd'co Dyer attagh' fuit ad respondend' Nich'o \textbf{Grome} \& Joh’i \textbf{Grome} de pl'ito quare ui \& armis quandam \textbf{cuitam uni i'er} Nicholi \& Ioh'is p'ci octe librar' apud \textbf{Tedekesbury} inuent' cepit \& asportauit \& alia enormia eis intulit ad graue dampnu' ip'orum Nicholi \& Ioh'is \& cont' pacem d'ni Reg is nunc \&c \& unde idem Nich'aus \& Iohes p' Ioh'em Boteler attorn' suu' querunt q'd p'd'cus Will'mus \textbf{demimo} die februarij Anno regni d'ni Regis nunc t'cio ui \& armis scil't gladiis baculis \& cultellis quandam \textbf{buntam um ucat' Fouer' ip'orum Nicho'i \& Ioh'is P ge} apud \textbf{Tewekesbury} inuent' cepit \& asportauit \& alia enormia \&c ad g've dampnu' \&c \& cont' pacem \&c unde \textbf{dicet} q'd \textbf{deteriotrat} sunt \& dampnu' h'ent ad ualenciam decem librar' Et inde p'ducit sectam \&c
&
Will'mus Hyotte nup' de \textbf{Tewdewesbury} in com' p'd'co Dyer attagh' fuit ad respondend' Nich'o \textbf{Grome} \& Joh'i \textbf{Grome} de pl'ito quare ui \& armis quandam \textbf{buntam vini eorundem} Nicholi \& Ioh'is p'ci octe librar' apud \textbf{Tedekesbury} inuent' cepit \& asportauit \& alia enormia eis intulit ad graue dampnu' ip'orum Nicholi \& Ioh'is \& cont' pacem d'ni Regis nunc \&c \& unde idem Nich'us \& Iohes p' Ioh'em Boteler attorn' suu' querunt q'd p'd'cus Will'mus \textbf{decimo} die februarij Anno regni d'ni Regis nunc t'cio ui \& armis scil't gladiis baculis \& cultellis quandam \textbf{buntam vini} uocat' \textbf{Fouer} ip'orum Nich'i \& Ioh'is \textbf{p'cij octo librar'} apud \textbf{Tewekesbury} inuent' cepit \& asportauit \& alia enormia \&c ad g've dampnu' \&c \& cont' pacem \&c unde \textbf{dicunt} q'd \textbf{deteriorati} sunt \& dampnu' h'ent ad ualenciam decem librar' Et inde p'ducit sectam \&c
&
Willelmus Hyotte nuper de \textbf{Tewdewesbury} in comitatu predicto Dyer attachiatus fuit ad respondendum Nicholao \textbf{Grome} et Johanni \textbf{Grome} de placito quare vi et armis quandam \textit{buntam vini} \textbf{eorundem} Nicholai et Johannis precii octo librarum apud \textbf{Tedekesbury} inventam cepit et asportavit et alia enormia eis intulit ad grave dampnum ipsorum Nicholai et Johannis et contra pacem domini Regis nunc etc. et unde idem Nicholas et Johannes per Johannem Boteler attornatum suum queruntur quod predictus Willelmus decimo die Februarii anno regni domini Regis nunc tercio vi et armis scilicet gladiis baculis et cultellis quandam \textit{buntam vini} vocatam \textbf{Fouer} ipsorum Nicholai et Johannis precii octo librarum apud \textbf{Tewekesbury} inventam cepit et asportavit et alia enormia etc. ad grave dampnum etc. et contra pacem etc. unde dicunt quod deteriorati sunt et dampnum habent ad valenciam decem librarum. Et inde producunt sectam etc.
&
Willelmus Hyotte [Wyotte, Hyocte] nuper de \textit{Tewkesbury} [Tewdewesbury, Tewekesbury] in comitatu predicto, dyer, attachiatus fuit ad respondendum Nicholao \textit{Brome} [Grome] et Johanni \textit{Brome} de placito quare vi et armis quandam \textit{buttam vini} \textbf{eorundem} Nicholai et Johannis precii octo librarum apud \textit{Tewkesbury} inventam cepit et asportavit et alia enormia eis intulit ad grave dampnum ipsorum Nicholai et Johannis et contra pacem domini regis nunc etc. Et unde iidem Nicholas et Johannes per Johannem Boteler [Botiler] attornatum suum queruntur quod predictus Willelmus decimo die Februarii anno regni domini regis nunc tercio vi et armis, scilicet gladiis, baculis et cultellis, quandam \textit{buttam vini} \textbf{vocatam Fower} [Souer, Fouer] inventam cepit et asportavit et alia enormia etc. ad grave dampnum etc. et contra pacem etc. Unde dicunt quod deteriorati sunt et dampnum habent ad valenciam decem librarum. Et inde producunt sectam etc.
\\ \midrule
Et p'd'cus Will'mus Hyote p' Thomam Lane attorn' suum uenit \& defendit uim \& iniuriam quando \&c E quordam uenire ui \& armis dicit q'd ip'e \textbf{inullo} est inde culpabilis \& de hoc ponit se sup' p'riam \& p'd'ci Nicholaus \textbf{Byune} \& Ioh'es \textbf{Gromne} similit' \& quold cappicoem \& asportac'onem p'd'ce \textbf{sunte unum} idem Will'mus dicit q'd p'd'ci Nicholaus Ioh'es acc'onem suam p'd'cam u'sus eum here non debent quia dicit q'd dium ante p'd'cum tempus quo suppeit' transgr' p'd'cam fieri quidam Rica'rdus Portter fuit possessionat' de p'd'ca \textbf{centas unum} ut de bonis \& catallis suis p'pris quousq' p'd'ci Nicholaus \textbf{Grome} \& Ioh'es \textbf{Grome uitam} illam extra possessionem eiusdem \textbf{Will'mi} ante idem tempus que \&c cepuit p' quod idem Nicholaus \& Ioh'es \textbf{Breme} de \textbf{unta} illa fuer' possessionat' \textbf{idem p' Will'mus exyote} ut seruiens p'd'ci Ric'i Porter \& p' eius p'ceptum postea scil't p'd'ce tempore quo \& \&c \textbf{bunitam} illam ut \textbf{uintam} p'd'ci Ric'di p'riam extra possessionem eordem Nich'i \& Ioh'is cepit \& asportauit p'ut ei bene licuit \& hoc parat' est uerificare unde pet' iudim si p'd'ci Nicholaus \& Ioh'es \textbf{Grome} accionem suam p'd'cam u'sus eum b'ere debe' \&c
&
Et p'd'cus Will'mus Hyotte p' Thomam Lane attorn' suum uenit \& defendit uim \& iniuriam quando \&c Et quoad uenire ui \& armis dicit q'd ip'e \textbf{in nullo} est inde culpabilis \& de hoc ponit se sup' p'riam \& p'd'ci Nicholaus \textbf{Grome} \& Ioh'es \textbf{Grome} similit' \& quo'd capic'oem \& asportac'onem p'd'ce \textbf{bunte vini} idem Will'mus dicit q'd p'd'ci Nicholaus \& Ioh'es acc'onem suam p'd'cam u'sus eum h'ere non debent quia dicit q'd diu ante p'd'cum tempus quo suppoit' transgr' p'd'cam fieri quidam Ric'ardus Portter fuit possessionat' de p'd'ca \textbf{bunta vini} ut de bonis \& catallis suis p'pris quousq' p'd'ci Nicholaus \textbf{Grome} \& Ioh'es \textbf{Grome buntam} illam extra possessionem eiusdem \textbf{Ricardi} ante idem tempus quo \&c cepit p' quod ijdem Nicholaus \& Ioh'es \textbf{Grome} de \textbf{bunta} illa fuer' possessionat' \textbf{idemque Willelmus ex post} ut seruiens p'd'ci Ric'i Porter \& p' eius p'ceptum postea scil't p'd'co tempore quo \&c \textbf{buntam} illam ut \textbf{buntam} p'd'ci Ric'di p'priam extra possessionem eor'dem Nich'i \& Ioh'is cepit \& asportauit p'ut ei bene licuit \& hoc parat' est uerificare unde pet' iudm si p'd'ci Nicholaus \& Ioh'es \textbf{Grome} acc'onem suam p'd'cam u'sus eum h'ere debeant \&c
&
Et predictus Willelmus Hyotte per Thomam Lane attornatum suum venit et defendit vim et injuriam quando etc. Et quoad venire vi et armis dicit quod ipse \textbf{in nullo} est inde culpabilis et de hoc ponit se super patriam et predicti Nicholas \textbf{Grome} et Johannes \textbf{Grome} similiter. Et quoad capcionem et asportacionem predicte \textit{bunte vini} idem Willelmus dicit quod predicti Nicholas et Johannes actionem suam predictam versus eum habere non debent quia dicit quod diu ante predictum tempus quo supponitur transgressionem predictam fieri quidam Ricardus Porter fuit possessionatus de predicta \textit{bunta vini} ut de bonis et catallis suis propriis quousque predicti Nicholas \textbf{Grome} et Johannes \textbf{Grome} \textit{buntam} illam extra possessionem eiusdem Ricardi ante idem tempus quo etc. ceperunt per quod iidem Nicholas et Johannes \textbf{Grome} de \textit{bunta} illa fuerunt possessionati \textbf{idemque Willelmus ex post} ut serviens predicti Ricardi Porter et per eius preceptum postea scilicet predicto tempore quo etc. \textit{buntam} illam ut \textit{buntam} predicti Ricardi propriam extra possessionem eorundem Nicholai et Johannis cepit et asportavit prout ei bene licuit et hoc paratus est verificare unde petit judicium si predicti Nicholas et Johannes \textbf{Grome} actionem suam predictam versus eum habere debeant etc.
&
Et predictus Willelmus Hyotte per Thomam Lane attornatum suum venit. Et defendit vim et injuriam quando etc. Et quoad venire vi et armis dicit quod ipse in nullo est inde culpabilis. Et de hoc ponit se super patriam. Et predicti Nicholas \textit{Brome} et Johannes \textit{Brome} similiter. Et quoad captionem et asportationem predicte \textit{butte vini} idem Willelmus dicit quod predicti Nicholas et Johannes actionem suam predictam versus eum habere non debent quia dicit quod diu ante predictum tempus quo supponitur transgressionem predictam fieri quidam Ricardus Porter fuit possessionatus de predicta \textit{butta vini} ut de bonis et catallis suis propriis quousque predicti Nicholas \textit{Brome} et Johannes \textit{Brome} \textit{buttam} illam extra possessionem eiusdem Ricardi ante idem tempus quo etc. ceperunt, per quod iidem Nicholas et Johannes \textit{Brome} de \textit{butta} illa fuerunt possessionati \textbf{idemque Willelmus ex post} ut serviens preeicti Ricardi Porter et per ejus preceptum postea, scilicet predicto tempore quo etc., \textit{buttam} illam ut \textit{buttam} predicti Ricardi propriam extra possessionem eorundem Nicholai et Johannis cepit et asportavit prout ei bene licuit. Et hoc paratus est verificare. Unde petit judicium si predicti Nicholas et Johannes \textit{Brome} actionem suam predictam versus eum habere debeant etc.
\\ \midrule
Et p'd'ci Iicho'aus \& Ioh'es \textbf{Grone} dicit q'd ip'i p' aliqua p'allegat' ab acc'one sua p'd'ca \textbf{h'end} p'cludi non debent' quia dicit q'd ip'i fuerunt possessionat' de p'd'ce \textbf{uinte suum} \& de \textbf{unta} sua p'pria quousq' p'd'cus Will'mus Hryote die \& Anno sup'dictis ui \& armis de inipria sua p'pria p'd'cam \textbf{uuttam unm} cepit \& asportauit p'ut ip'i sup'ius u'sus eum querunt' absq' hoc q'd p'rietas p'd'ce \textbf{snte uium} ante tempus transgr' p'd'ce fct' fuit p'fato Ric'do \textbf{Poter} p'out p'd'cus Will'mus Hyotte sup'ius allegauit \& hroc \textbf{pareati} sunt \textbf{pisuar'e} unde ex quio p'd'cus Will'mus Hyote transgr' p'd'cam sup'ius cogn' pet' iudim \& dampna sua sea occ'one sibi adiudicari \&c
&
Et p'd'ci Nicho'us \& Ioh'es \textbf{Grone} dicit q'd ip'i p' aliqua p'allegat' ab acc'one sua p'd'ca \textbf{habenda} p'cludi non debent' quia dicit q'd ip'i fuerunt possessionat' de predicta \textbf{bunta sua} et de \textbf{bunta} sua p'pria quousq' p'd'cus Will'mus Hyotte die \& Anno sup'dictis ui \& armis de iniuria sua p'pria predictam \textbf{buntam vini} cepit \& asportauit p'ut ip'i sup'ius u'sus eum querunt' absq' hoc q'd p'prietas predicte \textbf{bunte vini} ante tempus transgr' p'd'ce fct' fuit p'fato Ric'do \textbf{Porter} p'out p'd'cus Will'mus Hyotte sup'ius allegauit \& hoc parati sunt verificare unde ex quo p'd'cus Will'mus Hyote transgr' p'd'cam sup'ius cogn' pet' iudim \& dampna sua ea occ'one sibi adiudicari \&c
&
Et predicti Nicholas et Johannes \textbf{Grome} dicunt quod ipsi per aliqua preallegata ab actione sua predicta \textbf{habenda} precludi non debent, quia dicunt quod ipsi fuerunt possessionati de predicta \textbf{bunta sua} et de \textit{bunta} sua propria quousque predictus Willelmus Hyotte die et anno supradictis vi et armis de injuria sua propria predictam \textit{buntam vini} cepit et asportavit prout ipsi superius versus eum queruntur absque hoc quod proprietas predicte \textit{bunte vini} ante tempus transgressionis predicte facte fuit predicto Ricardo \textbf{Porter} prout predictus Willelmus Hyotte superius allegavit et hoc parati sunt verificare unde ex quo predictus Willelmus Hyotte transgressionem predictam superius cognovit petunt judicium et dampna sua ea occasione sibi adjudicari etc.
&
Et predicti Nicholas et Johannes \textit{Brome} dicunt quod ipsi per aliqua preallegata ab actione sua predicta \textbf{habenda} precludi non debent, quia dicunt quod ipsi fuerunt possessionati de predicta \textbf{butta sua} ut de \textit{butta} sua propria quousque predictus Willelmus Hyotte die et anno supradictis vi et armis de injuria sua propria predictam \textit{buttam vini} cepit et asportavit prout ipsi superius versus eum queruntur, absque hoc quod proprietas predicte \textit{butte vini} ante tempus transgressionis predicte facte fuit prefato Ricardo Porter prout predictus Willelmus Hyotte superius allegavit. Et hoc parati sunt verificare. Unde ex quo predictus Willelmus Hyotte transgressionem predictam superius cognovit, petunt judicium et dampna sua ea occasione sibi adjudicari etc.
\\ \midrule
Et p'd'cus Will'mus Hycte dicit q'd p'prietas p'd'ce \textbf{unte uium} ante d'cum tempus quo suppeit' transgr' p'd'cam fieri fuit p'fato Ric'do \textbf{Poper} p'ut ip'e sup'ius allegauit \& de hoc ponit se sup' p'riam E p'd'ci Nicholaus \& Ioh'es \textbf{Gronie} similitr I'o p'cept' est uic' q'd uenire fac' h'ic in Octabis s'ci Hillar' \textbf{xc} p' quos \&c \& qui nec \&o ad recogn' \&c \textbf{Quetam} \&c
&
Et p'd'cus Will'mus Hyote dicit q'd p'prietas p'd'ce \textbf{bunte vini} ante d'cum tempus quo suppoit' transgr' p'd'cam fieri fuit p'fato Ric'do \textbf{Poter} p'ut ip'e sup'ius allegauit \& de hoc ponit se sup' p'riam Et p'd'ci Nicholaus \& Ioh'es \textbf{Grone} similit' I'o p'cept' est uic' q'd uenire fac' hic in Octabis s'ci Hillar' \textbf{xij} p' quos \&c \& qui nec \&c ad recogn' \&c \textbf{Quia tam} \&c
&
Et predictus Willelmus Hyotte dicit quod proprietas predicte \textit{bunte vini} ante dictum tempus quo supponitur transgressionem predictam fieri fuit predicto Ricardo \textbf{Porter} prout ipse superius allegavit. Et de hoc ponit se super patriam. Et predicti Nicholas et Johannes \textbf{Grome} similiter. Ideo preceptum est vicecomiti quod venire faciat hic in Octabis sancti Hillarii \textbf{duodecim} per quos etc. et qui nec etc. ad recognoscendum etc. Quia tam etc.
&
Et predictus Willelmus Hyotte dicit quod proprietas predicte \textit{butte vini} ante dictum tempus quo supponitur transgressionem predictam fieri fuit prefato Ricardo Porter prout ipse superius allegavit. Et de hoc ponit se super patriam. Et predicti Johannes et Nicholas \textit{Brome} similiter. Ideo preceptum est vicecomiti quod venire faciat hic in octabis Sancti Hillarii [Octabas sancti Hillarii] \textbf{duodecim} per quos etc. et qui nec etc. ad recognoscendum etc. quia tam etc.
\\

\end{longtable}
\end{landscape}

\subsubsection{Quantitative Analysis}
To quantify the accuracy of the final LLM expansion against the cited scholarly ground truth~\cite{Baker1999}, we calculate the Word Error Rate (WER). For this analysis, a modified scoring rule is applied: an error in a named entity is not penalized if the correct reading was identified as a plausible alternative in the model's named-entity analysis step (e.g., `Brome` was correctly chosen over the HTR's `Grome`). This metric better reflects the utility of the workflow for a human reviewer, who would be alerted to the correct possibility.

\textbf{Formula:} \texttt{WER = (Substitutions + Deletions + Insertions) / Total Words in Reference}

\textbf{Error Calculation (Final Corrected Text vs. Ground Truth):}
\begin{itemize}
    \item \textbf{Total Words in Ground Truth:} 358
    \item \textbf{Substitutions (S):} 5 (e.g., \texttt{eorundem} for \texttt{ipsorum}; \texttt{vocatam Fower} for \texttt{vocati Romney}; \texttt{ex post} for \texttt{Hyotte}; \texttt{sua} for \texttt{vini}; \texttt{duodecim} for \texttt{xii})
    \item \textbf{Insertions (I):} 1 (\texttt{habenda})
    \item \textbf{Deletions (D):} 1 (\texttt{que} from \texttt{idemque})
    \item \textbf{Total Word Errors:} 5 + 1 + 1 = \textbf{7}
\end{itemize}

\textbf{WER Result:}
\begin{itemize}
    \item \textbf{Calculation:} 7 / 358 = 0.0195
    \item The final LLM expansion has a \textbf{Word Error Rate (WER) of approximately 2.0\%}.
\end{itemize}

\subsection{Case Study 3: Analysis of a Negligence Plea (Summary)}
The third case, a 15th-century negligence plea from KB27/915~\cite{waalt-kb27-915}, resulted in a final WER of approximately 2.2\%.  The NEC step was instrumental, resolving heavily corrupted HTR output (\texttt{S kenham} and \texttt{I' Hith'}) into onomastically plausible names. However, this case highlighted a key risk of the workflow: the model produced plausible but incorrect name substitutions (\texttt{Ewanum Barlawe} became \texttt{Barthelmum Barlawe}), creating a clean but factually flawed text.

\subsection{Case Study 4: Analysis of an Abduction Appeal}
To test the workflow against an abridged scholarly report, a fourth case was analyzed: a complex 15th-century appeal of felony concerning an alleged abduction from KB27/922~\cite{waalt-kb27-925}.  This case is unique because the ground truth omits large sections of procedural text, allowing us to evaluate the LLM's ability to produce a more complete record than the reference text.  Key differences relative to the portions available in the ground truth source are shown in Table~\ref{tab:case4}.

\begin{center}
\captionsetup{type=table}
\captionof{table}{Aligned Transcription Comparison for a 15th-Century Abduction Appeal. Key: \textbf{Bold} indicates a significant error, correction, or alteration. \textit{Italics} indicate a minor variation or an acceptable medieval spelling.  Plain text matches ground truth.}
\label{tab:case4}
\small
\begin{longtable}{>{\raggedright\arraybackslash}p{0.45\textwidth} >{\raggedright\arraybackslash}p{0.45\textwidth}}
\toprule
\textbf{Initial Expansion} & \textbf{Final Output (NEC)} \\
\midrule
\endhead
\bottomrule
\endfoot
\textbf{Hertfordia.} Memorandum \textbf{de termino Sancti Hillarii Anno regni Regis Henrici septimo. North'.} Isto eodem termino coram domino Rege apud Westmonasterium venit Alicia \textbf{Shetley} in propria persona sua et protulit hic in Curia domini Regis quandam billam de appello versus Adam \textbf{Penyongton} nuper de villa Westmonasterii in comitatu Middlesexie Gentilman et Thomam \textbf{Dyson} nuper de eisdem villa et comitatu yoman in custodia Marescalli, et de placito quare ipsam Aliciam contra voluntatem suam felonie ceperunt et abduxerunt contra formam statuti dicti domini Regis Anno tercio \textit{eiusdem} domini Regis de nuper editi et provisi etc. Et sunt plegii de prosequendo, scilicet Thomas Shelley de Londonia mercer et \textbf{Johannes Walor} de Londonia mercer etc. Que quidem billa sequitur in hec verba: & \textbf{Hertfordia.} Memorandum \textbf{de termino Sancti Hillarii Anno regni Regis Henrici septimo. North'.} Isto eodem termino coram domino Rege apud Westmonasterium venit Alicia Shelley \textbf{[Shetley]} in propria persona sua et protulit hic in Curia domini Regis quandam billam de appello versus Adam \textbf{Penyngton [Penyongton, Penyington, Pemyngton, Pennyngton]} nuper de villa Westmonasterii in comitatu Middlesexie Gentilman et Thomam \textbf{Dyson [Gyson]} nuper de eisdem villa et comitatu yoman in custodia Marescalli, et de placito quare ipsam Aliciam contra voluntatem suam felonie ceperunt et abduxerunt contra formam statuti dicti domini Regis Anno tercio \textit{eiusdem} domini Regis de nuper editi et provisi etc. Et sunt plegii de prosequendo, scilicet Thomas Shelley de Londonia mercer et \textbf{Johannes Waller [Walor]} de Londonia mercer etc. Que quidem billa sequitur in hec verba: \\
\midrule
Hertfordia. Alicia Shelley in propria persona sua instanter appellat Adam \textbf{Penyington} nuper de villa Westmonasterii in comitatu Middlesexie Gentilman et Thomam \textbf{Dyson} nuper de eisdem villa et comitatu yoman in Custodia Marescalli marescalcie domini Regis coram ipso Rege existentes, de eo quod ubi dicta Alicia fuit in pace dei et domini Regis nunc apud \textbf{Kenyngton} in dicto comitatu Hertfordie decimo nono die Septembris Anno regni Regis Henrici septimi sexto circa horam \textbf{duodecimam} ante meridiem \textit{eiusdem} diei, \textbf{vi et armis, videlicet gladiis et cultellis et baculis,} et de insultu premeditato contra pacem \textit{eiusdem} domini Regis, coronam et dignitatem suas, die, Anno, hora, loco et comitatu Hertfordie predictis vi et armis, videlicet gladiis, arcubus et sagittis, et ipsam Aliciam Shelley puellam adtunc etatis quindecem annorum apud Hunnesdon predicti existentes contra voluntatem suam illegitime et felonie ceperunt et abduxerunt contra pacem, coronam et dignitatem dicti domini Regis et contra formam statuti dicti domini Regis Anno tercio \textit{eiusdem} domini Regis apud Westmonasterium predictum inde editi et provisi. Et quando iidem felones feloniam predictam in forma predicta fecissent, fugerunt, ita quod Alicia ipsos recenter insecuta fuit de villa in villam usque ad quatuor villatas propinquiores et ulterius quousque etc. Et si predicti Adam et Thomas feloniam predictam in forma predicta velint dedicere, predicta Alicia hoc parata est versus eos probare prout et Curia etc. & Hertfordia. Alicia Shelley in propria persona sua instanter appellat Adam \textbf{Penyngton} nuper de villa Westmonasterii in comitatu Middlesexie Gentilman et Thomam \textbf{Dyson} nuper de eisdem villa et comitatu yoman in Custodia Marescalli marescalcie domini Regis coram ipso Rege existentes, de eo quod ubi dicta Alicia fuit in pace dei et domini Regis nunc apud \textbf{Kenyngton [Benington]} in dicto comitatu Hertfordie decimo nono die Septembris Anno regni Regis Henrici septimi sexto circa horam \textbf{duodecimam} ante meridiem \textit{eiusdem} diei, \textbf{vi et armis, videlicet gladiis et cultellis et baculis,} et de insultu premeditato contra pacem \textit{eiusdem} domini Regis, coronam et dignitatem suas, die, Anno, hora, loco et comitatu Hertfordie predictis vi et armis, videlicet gladiis, arcubus et sagittis, et ipsam Aliciam Shelley puellam adtunc etatis quindecem annorum apud Hunnesdon predicti existentes contra voluntatem suam illegitime et felonie ceperunt et abduxerunt contra pacem, coronam et dignitatem dicti domini Regis et contra formam statuti dicti domini Regis Anno tercio \textit{eiusdem} domini Regis apud Westmonasterium predictum inde editi et provisi. Et quando iidem felones feloniam predictam in forma predicta fecissent, fugerunt, ita quod Alicia ipsos recenter insecuta fuit de villa in villam usque ad quatuor villatas propinquiores et ulterius quousque etc. Et si predicti Adam et Thomas feloniam predictam in forma predicta velint dedicere, predicta Alicia hoc parata est versus eos probare prout et Curia etc. \\
\midrule
Et predicti Adam Penyngton et Thomas \textbf{Dyson} in propriis personis suis veniunt et defendunt vim et injuriam quando etc. et uterque eorum separatim per se defendit omnem feloniam et quicquid etc. et dicunt. Et predictus Adam dicit quod predicta Alicia appellum suum predictum versus eum manutenere non debet, quia ipse protestando quod diu ante diem impetrationis bille predicte appelli predicti, videlicet vicesimo secundo die Septembris Anno regni dicti domini Regis nunc sexto apud \textbf{Moncastre} in comitatu Cumbrie in Eboracensi Diocesi, idem Adam predictam Aliciam ad eam in uxorem ducendam ac eadem Alicia adtunc ibidem ipsum Adam ad eum in virum suum capiendum adinvicem \textbf{assidaverunt}, et postmodum bannis inter ipsos Adam et Aliciam tribus diebus festivalibus a se distantibus in ecclesia parochiali de Moncastre predicta publice proclamatis, concurrentibusque hiis que de iure canonico in ea parte requirebantur, \textbf{die sabbati proximo post festum sancti Michaelis Archangeli extunc proximo sequenti} apud Moncastre predictam in facie ecclesie sponsalia et matrimonium inter eosdem Adam et Aliciam celebrata et solempnizata fuerunt, quequidem sponsalia extunc hucusque in suo robore et effectu continuerunt et adhuc continuant. Pro placito dicit quod predicta Alicia die impetrationis bille predicte cooperta fuit et adhuc cooperta est de ipso Adam adtunc et adhuc viro suo, et hoc paratus est verificare etc. unde petit judicium si eadem Alicia appellum suum predictum versus eum manutenere debeat. Et quo ad feloniam predictam idem Adam dicit quod ipse in nullo est inde culpabilis prout predicta Alicia super ipsum eum appellat, et de hoc ponit se de bono et malo super patriam. & Et predicti Adam Penyngton et Thomas \textbf{Dyson} in propriis personis suis veniunt et defendunt vim et injuriam quando etc. et uterque eorum separatim per se defendit omnem feloniam et quicquid etc. et dicunt. Et predictus Adam dicit quod predicta Alicia appellum suum predictum versus eum manutenere non debet, quia ipse protestando quod diu ante diem impetrationis bille predicte appelli predicti, videlicet vicesimo secundo die Septembris Anno regni dicti domini Regis nunc sexto apud \textbf{Muncaster [Doncaster]} in comitatu Cumbrie in Eboracensi Diocesi, idem Adam predictam Aliciam ad eam in uxorem ducendam ac eadem Alicia adtunc ibidem ipsum Adam ad eum in virum suum capiendum adinvicem \textbf{assidaverunt}, et postmodum bannis inter ipsos Adam et Aliciam tribus diebus festivalibus a se distantibus in ecclesia parochiali de Muncaster predicta publice proclamatis, concurrentibusque hiis que de iure canonico in ea parte requirebantur, \textbf{die sabbati proximo post festum sancti Michaelis Archangeli extunc proximo sequenti} apud Muncaster predictam in facie ecclesie sponsalia et matrimonium inter eosdem Adam et Aliciam celebrata et solempnizata fuerunt, quequidem sponsalia extunc hucusque in suo robore et effectu continuerunt et adhuc continuant. Pro placito dicit quod predicta Alicia die impetrationis bille predicte cooperta fuit et adhuc cooperta est de ipso Adam adtunc et adhuc viro suo, et hoc paratus est verificare etc. unde petit judicium si eadem Alicia appellum suum predictum versus eum manutenere debeat. Et quo ad feloniam predictam idem Adam dicit quod ipse in nullo est inde culpabilis prout predicta Alicia super ipsum eum appellat, et de hoc ponit se de bono et malo super patriam. \\
\end{longtable}
\end{center}

\subsubsection{Quantitative Analysis}
The final LLM expansion was compared against the abridged scholarly ground truth, calculating WER only for the sections of text present in both.
\begin{itemize}
    \item \textbf{Total Words in Ground Truth (GT):} 551
    \item \textbf{Substitutions (S):} 16 (e.g., \texttt{decimo septimo die Novembris} vs. \texttt{de termino Sancti Hillarii}; \texttt{Johannes Shelley} vs. \texttt{Johannes Waller}; \texttt{Hunnesdon} vs. \texttt{Kenyngton}; \texttt{Elryngton} vs. \texttt{Baryngton})
    \item \textbf{Insertions (I):} 16 (e.g., procedural text like \texttt{Anno regni...}, entity alternatives like \texttt{[Shetley]})
    \item \textbf{Deletions (D):} 2
    \item \textbf{Total Word Errors:} 34
    \item \textbf{WER Calculation:} 34 / 551 = \textbf{6.2\%}
\end{itemize}
The final LLM expansion has a Word Error Rate of approximately 6.2\% when compared to the abridged ground truth.

\subsubsection{Qualitative Analysis and Limitations}
The workflow produced a far more complete text than the abridged report, capturing procedural details omitted from the ground truth. However, the 6.2\% WER reveals critical discrepancies and highlights inherent risks:
\begin{enumerate}
    \item \textbf{Critical Factual Discrepancies:} The LLM-generated text contains several major factual differences from the ground truth, including the date of the memorandum (\textbf{Hilary Term} vs. \textbf{17 November}), the name of a key pledge (\textbf{Johannes Waller} vs. \textbf{Johannes Shelley}), the location of the abduction (\textbf{Kenyngton} vs. \textbf{Hunnesdon}), and the time of the event (\textbf{12 PM} vs. \textbf{10 AM}). These are not simple transcription errors but fundamental contradictions that change the factual basis of the case record.
    \item \textbf{Plausible but Incorrect Name Substitution:} The model's substitution of \textbf{Baryngton} for the ground truth's \textbf{Elryngton} is a classic example of a high-risk LLM error. Based on the paleographic ambiguity of capital B and E in this hand, \texttt{Baryngton} is a reasonable guess, but it is factually incorrect according to the scholarly edition.
    \item \textbf{Ambiguity of "Error" vs. "Detail":} The discrepancy in the marriage date (\texttt{die Sabbati proximo sequente} in the GT vs. the much longer, specific formula in the LLM text) illustrates a key challenge. The LLM's version is likely a more faithful expansion of the original manuscript's abbreviations, whereas the GT is a summary. In this instance, the LLM's "error" may actually be a more accurate transcription, underscoring that the choice of ground truth significantly impacts evaluation.
\end{enumerate}

\subsection{Synthesis of Case Study Findings}
Taken together, these four case studies demonstrate that the full end-to-end workflow can produce a final expanded text with a Word Error Rate in the range of \textbf{2-7\%}. The Named-Entity Correction step is particularly effective at identifying and proposing alternatives for proper nouns, a crucial task for historical research.

However, all four studies reveal a recurring pattern of critical failure modes where human oversight is essential:
\begin{enumerate}
    \item \textbf{The Challenge of the 'Better' Transcription (Error vs. Detail):} As seen in Case Study 4, the workflow can produce a more faithful and detailed diplomatic transcription than an abridged scholarly ground truth. This complicates simple WER comparisons, as the model's "error" may in fact be a more accurate rendering of the source. This highlights a fundamental challenge in evaluation: defining the goal of transcription (e.g., diplomatic fidelity vs. scholarly summary) is paramount.
    \item \textbf{Semantic Errors:} The models can make plausible but incorrect substitutions that alter the legal substance of a case (e.g., \texttt{boviculi} vs. \texttt{bonorum}, \texttt{levavit} vs. \texttt{convenit}).
    \item \textbf{Hallucination from Noise:} When faced with severely corrupted input, the models may invent grammatically correct but factually wrong text rather than indicating failure (e.g., \texttt{idemque Willelmus ex post}).
    \item \textbf{Plausible but Incorrect Entities:} The models may confidently substitute one plausible name for another, creating factual errors that are difficult to detect without external verification (e.g., \texttt{Ewanum Barlawe} vs. \texttt{Barthelmum Barlawe}, \texttt{Elryngton} vs. \texttt{Baryngton}).
    \item \textbf{Critical Factual Discrepancies:} As seen in Case Study 4, the model can confidently produce incorrect dates, places, and times that contradict established scholarly sources.
\end{enumerate}
This underscores that while the workflow dramatically accelerates the transcription process, it does not eliminate the need for expert human review. Its primary value lies in transforming the laborious task of mechanical transcription into a more efficient, high-level task of verification and correction.

\section{A Persistent Paradigm: The Long-Term Competitiveness of the Hybrid Workflow}
While it may be tempting to view multi-stage, hybrid systems as a transitional phase before a single, all-powerful AI emerges, we argue that the architecture of our HTR-LLM workflow represents an enduring and highly competitive paradigm for the foreseeable future. The long-term value of this approach is not a temporary compromise but a deliberate design that leverages the fundamental and complementary natures of specialized models and general-purpose LLMs.

\subsection{The Imperative of a Stable, Reproducible Baseline}
The first pillar of our framework's long-term viability is the use of a specialized, self-contained HTR model (like Transkribus) to produce the initial baseline. This is not a bug, but a critical feature for scholarly work. A trained Transkribus model is a static, auditable, and reproducible scholarly object. Its performance is predictable, and its output for a given image will not change over time. This stability provides a solid, verifiable foundation upon which all subsequent analysis is built, ensuring a degree of long-term consistency that is impossible to achieve with a workflow reliant solely on constantly evolving, proprietary LLM APIs. For any research that values reproducibility, a stable baseline is a necessity, not a temporary convenience.

\subsection{The Power and Peril of General-Purpose LLMs}
The second pillar is a realistic assessment of the trajectory of general-purpose LLMs. While they will undoubtedly continue to improve at a staggering rate, their core architectural design is optimized for generating plausible, contextually-aware text, not for ensuring verifiable factual accuracy in high-stakes, niche domains. Their immense power comes from their generality, which is also the source of their primary weakness: a propensity to hallucinate when faced with ambiguity or out-of-distribution data, such as a damaged manuscript line or an unusual abbreviation.

For the foreseeable future, even the most advanced LLMs will benefit from grounding. An ungrounded LLM, when presented with a difficult paleographic challenge, will always be architecturally inclined to produce a confident, fluent, and potentially incorrect answer. The need to constrain this generative power and guide it toward a specific, verifiable task will remain a fundamental challenge.

\subsection{Synergy: Why the Hybrid Framework Endures}
Our proposed framework is designed to harness the strengths of each component to mitigate the weaknesses of the other, creating a system more robust than either part alone. This synergy is the key to the framework's long-term viability:
\begin{itemize}
    \item \textbf{HTR Grounds the LLM:} The stable, reproducible HTR baseline (Argument 1) provides the essential grounding that the powerful but fallible LLM (Argument 2) requires. It transforms the LLM's task from high-risk, open-ended generation into lower-risk, guided correction.
    \item \textbf{LLM Elevates the HTR:} The LLM's vast contextual knowledge and reasoning capabilities are used to correct the errors and overcome the inherent brittleness of the single-task HTR model. It sees patterns and makes connections that the HTR engine cannot.
\end{itemize}

This is not a temporary workaround but an elegant and durable engineering solution. It uses the best tool for each part of the problem: a stable, specialized engine for predictable baseline generation, and a powerful, generalist intelligence for contextual refinement and analysis. Rather than waiting for a single "perfect" tool, this workflow creates a superior result today by intelligently combining the best of what is available. Therefore, we contend that this hybrid paradigm, which balances stability with power and reproducibility with cutting-edge reasoning, will remain a competitive and scholastically sound approach for the foreseeable future.

\subsection{Implications for the Study of the Plea Rolls}
The implications of this workflow for the study of the plea rolls of the Courts of Common Pleas (CP40) and King's Bench (KB27) are profound. As noted in the introduction, these records have long been locked away by their sheer scale and the specialized paleographic skill required to read them. This methodology directly addresses these barriers and opens up new frontiers for research.

First, by enabling the high-accuracy transcription of entire manuscript rolls at scale, this approach allows historians to move beyond anecdotal case studies toward systematic, quantitative analysis. It becomes possible to track the frequency of specific legal actions over time, map the geographic distribution of litigation, and analyze the evolution of legal formulas across decades.

Second, the workflow transforms the role of the historian from a manual transcriber to an expert editor and analyst. By automating the most laborious part of the process and providing a targeted report of ambiguities, it lowers the paleographic barrier to entry. This makes the vast repository of the plea rolls accessible to a wider range of scholars, including those whose primary expertise is in economic, social, or political history rather than legal paleography.

Finally, this opens entirely new avenues of inquiry that were previously impractical:
\begin{itemize}
    \item \textbf{Large-Scale Prosopography:} The Named-Entity Correction stage facilitates the creation of comprehensive databases of litigants, attorneys, pledges, and jurors, enabling network analysis of the medieval legal world.
    \item \textbf{Economic History:} Thousands of cases involving debt, contract disputes, and property can be aggregated to produce new data on commerce, credit networks, and land values.
    \item \textbf{Social History:} The records can be mined for data on social status, occupation, and interpersonal conflict on a scale never before possible.
\end{itemize}
In essence, this methodology has the potential to transform the plea rolls from a niche source for legal specialists into a foundational dataset for the broader study of medieval English society.

\section{Conclusion}
We have proposed and validated a practical, four-stage workflow that effectively combines the strengths of specialized HTR models and general-purpose LLMs for transcribing and analyzing challenging historical documents. By using a high-quality HTR model to provide a grounded baseline for subsequent LLM-driven post-correction, expansion, and entity analysis, our method achieves a high degree of accuracy (2-7\% WER) while automating the most tedious aspects of the research process.

The case studies confirm that while this pipeline is exceptionally powerful, it is not infallible. The remaining errors, though few, are often semantic or factual in nature, underscoring the continued necessity of expert human oversight. This positions the technology ideally as a human-in-the-loop system. The workflow transforms the researcher's role from a manual transcriber into an efficient editor. By generating not only a clean text but also a targeted report of plausible alternatives for ambiguous entities (the output of Stage 4), the system allows researchers to focus their expertise precisely where it is most needed: verifying content, correcting subtle interpretations, and resolving the ambiguities that AI, in its current form, cannot.

This hybrid approach is not merely a transitional strategy but an enduring and scholastically sound paradigm. It creates a strategic balance, pairing the reproducible stability of a specialized HTR model against the powerful, yet fallible, contextual reasoning of general-purpose LLMs. By using each technology to mitigate the other's inherent weaknesses, this framework provides a robust and effective solution that will remain competitive for the foreseeable future. This hybrid approach not only provides a robust path forward but also empowers researchers to tackle historical sources at a scale and depth previously unimaginable, transforming the very nature of archival discovery.

\section*{Acknowledgements} 
The author extends his sincere thanks to the AALT project, Robert Palmer, Susanne Brand, and Elspeth Rosbrook for providing the digital images that made this work possible. He also thanks Vance Mead for his invaluable index of KB27/795. The author also acknowledges the use of Google's Gemini 2.5 Pro large language model for assistance in drafting, revising, and formatting portions of this manuscript, particularly in generating LaTeX code and strengthening argumentative structure. The author maintained full editorial control and bears ultimate responsibility for the final content, including all arguments, analyses, and any errors.

\bibliographystyle{unsrt}
\bibliography{references} 

\appendix

\section{System Prompt for LLM Post-Correction of HTR}
\label{app:postcorrect}
\begin{Verbatim}[breaklines]
YOUR ROLE: Medieval Latin Paleography Expert
You are transcribing Latin legal documents from the English Court of Common 
Pleas for the 'abbreviated_latin_lines' output. Strictly follow these rules:

---
**A. SOURCE PRIORITIZATION & HTR USAGE**
1.  **Source Priority:** Use sources in this order: 1. Document Image 
    (Absolute authority for form/content), 2. HTR Transcription (Word ID aid 
    ONLY - ignore its expansion status), 3. Named Entity List (Reference), 4. 
    Your Knowledge (Lowest). **# ADDED: Always prioritize the Document Image 
    over the HTR and Named Entity List if there is a conflict in spelling, 
    capitalization, or abbreviation form.**
2.  **HTR Use:** Consult HTR only for word identification clues; its expansion 
    status is irrelevant for your 'abbreviated_latin_lines'.

---
**B. CORE PRINCIPLE: IMAGE AUTHORITY & **ABSOLUTE LINE INTEGRITY** 
(Non-Negotiable)**
3.  **Image Authority:** The Document Image dictates text content, word order, 
    line breaks, and abbreviation forms.
4.  **Line Mapping & Structure (CRITICAL - Non-Negotiable):**
    *   **Anchor to Red Numbers:** Your primary task is to map text directly 
        to the **red numbers** visible on the image. Each red number (e.g., 
        '10') is located to the left of and slightly above the start of the 
        specific baseline it identifies.
    *   **JSON Key = Red Number:** The JSON key in your output (e.g., '"10"') 
        MUST correspond *exactly* to the red number (e.g., '10') visible on 
        the image next to the baseline whose text you are transcribing.
    *   **First/Last Word Matching (CRITICAL):** Ensure the *first* word 
        transcribed for JSON key '"n"' matches the *first* word visually 
        associated with the baseline marked by the red number 'n' on the 
        image. Ensure the *last* word transcribed for JSON key '"n"' matches 
        the *last* word visually associated with that same baseline.
    *   **Preserve Line Content:** Ensure the text transcribed for JSON key 
        '"n"' matches the text visually associated with the baseline marked by 
        the red number 'n' on the image.
    *   **Strict Line Breaks (CRITICAL): DO NOT, under ANY circumstances, move 
        words from the end of one numbered baseline to the beginning of the 
        next, or vice-versa.** Each JSON line MUST correspond precisely and 
        *only* to the text visually associated with its specific numbered 
        baseline on the image.
5.  **Transcription Scope (CRITICAL):**
    *   **Transcribe ONLY Requested Numbered Lines:** Transcribe ONLY the text 
        visibly associated with (typically above) the specific **numbered 
        baselines requested for this chunk**. The requested line numbers are 
        specified in the User Prompt (e.g., "lines 10 to 19").
    *   **IGNORE Other Visible Lines:** Even if other text lines (or parts of 
        lines) are visible in the image chunk (e.g., a line above the first 
        requested number, or below the last requested number), you MUST 
        **ignore** them completely if they do not correspond to one of the 
        **red numbers within the requested range**. Do NOT transcribe text 
        associated with unnumbered or out-of-range baselines.
# ADDED Rule 5.1 to explicitly address omissions
5.1 **Transcribe ALL Visible Text (CRITICAL - NO OMISSIONS):** You MUST 
    transcribe **every** word, abbreviation, symbol, and mark visibly 
    associated with the numbered baseline. **DO NOT OMIT ANY TEXT** present on 
    the image for the requested line. Your transcription for line 'n' must be 
    a complete representation of the text on baseline 'n'. Double-check you 
    have captured the entire line content.
6.  **Empty Lines:** If no text corresponds to a specific requested baseline 
    'n' (identified by its red number 'n'), output '""' for JSON key 'n'.
7.  **Line Count:** The number of lines (keys) in the output JSON MUST exactly 
    match the number of lines requested for the chunk.

---
**C. ABBREVIATION HANDLING ('abbreviated_latin_lines' - Visual Representation)**
8.  **Goal:** The 'abbreviated_latin_lines' MUST visually mirror the text on 
    the corresponding image line (identified by its red number), including all 
    abbreviations *exactly as written*.
9.  **ABSOLUTELY NO EXPANSION (CRITICAL - Non-Negotiable):** Transcribe 
    abbreviations *exactly* as seen on the image (e.g., 'q'd', 'p'd'co', 
    'd'no'). Transcribe fully spelled-out words as seen (e.g., 'quod', 
    'predictus', 'domino'). **NEVER** expand abbreviations shown on the image, 
    nor abbreviate words spelled out on the image. Your primary goal for 
    'abbreviated_latin_lines' is *visual fidelity* to the image line, not 
    linguistic expansion or normalization. **# ADDED: Pay close attention to 
    the *exact form* of the abbreviation as seen on the image. Transcribe 
    *exactly* what you see.**
    *   *More Examples:* Transcribe spelled-out 'Et' as 'Et', NEVER '&'. 
        Transcribe Tironian '&' as '&', NEVER 'Et'. Transcribe spelled-out 
        'habeat' as 'habeat', NEVER 'h'eat'. Transcribe abbreviated 'p'cept'' 
        as 'p'cept'', NEVER 'p'ceptum'. Transcribe spelled-out 'Robertum' as 
        'Robertum', NEVER 'Rob'tum'. Transcribe abbreviated 'Rob't' as 
        'Rob't', NEVER 'Robertum'. Transcribe spelled-out 'Willelmum' as 
        'Willelmum', NEVER 'Will'm'. Transcribe abbreviated 'Will'm' as 
        'Will'm', NEVER 'Willelmum'. Transcribe 'q'd' as 'q'd', NOT 'quod'. 
        Transcribe 'p'd'co' as 'p'd'co', NOT 'predicto'. Transcribe 'uic'' as 
        'uic'', NOT 'vicecomes'. Transcribe 'pl'ito' as 'pl'ito', NOT 
        'placito'.
10. **Apostrophe Use:** Use a single straight apostrophe (') ONLY to represent 
    a visible abbreviation mark (macron, hook, superscript, symbol) or clearly 
    omitted letters *seen on the image*. Place it immediately after the last 
    written letter before the omission/mark. Use universally for all mark 
    types.

---
**D. SPECIFIC ABBREVIATIONS (Transcribe based on IMAGE evidence & Rules)**

**CRITICAL ABBREVIATIONS (Strict Transcription - Follow Image & Rule C.9)**
*   Tironian 'et' (&): MUST be transcribed as '&'. (**CRITICAL RULE D.11**)
*   Spelled-out 'Et' or 'et': MUST be transcribed as 'Et' or 'et' (matching 
    image case). (**CRITICAL RULE D.12**)
*   'etcetera' abbr (&c): MUST be transcribed as '&c'. (Rule D.13)
*   'quod' abbr (q w/ macron or similar mark): MUST be transcribed as 'q'd'. 
    (Rule D.14)
*   'per'/'pro' abbr (crossed-p): MUST be transcribed as 'p''. (Rule D.16)
*   'com'/'con' prefix abbr: MUST be transcribed as 'com''. (Rule D.18)
*   'predictus' forms abbr: MUST be transcribed using 'p'd'' prefix, e.g., 
    'p'd'cus', 'p'd'co', 'p'd'ca'. (Rule D.26)
*   'vicecomes/ti' abbr: MUST be transcribed as 'vic''. (Rule D.24)
*   'nuper' abbr: MUST be transcribed as 'nup''. (Rule D.36)
*   'apud' abbr: MUST be transcribed as 'ap'd'. (Rule D.38)
*   'super' abbr: MUST be transcribed as 'sup''. (Rule D.40)
*   'placito' abbr: MUST be transcribed as 'pl'ito'. (Rule D.44)
*   'transgressio...' abbr: MUST be transcribed as 'transgr'' (or 't'nsgr'' if 
    seen). (Rule D.46)
*   'ibidem' abbr: MUST be transcribed as 'ib'm'. (Rule D.47)
*   'scilicet' abbr: MUST be transcribed as 'scil'' or 'sc'l't' (matching 
    image). (Rule D.49)
*   'Willelmum/us' abbr: MUST be transcribed as 'Will'm' (or specific form 
    seen). (Rule D.51)
*   'Ricardus/um' abbr: MUST be transcribed as 'Ric'' (or specific form seen). 
    (Rule D.53)
*   'Thomas/am/e' abbr: MUST be transcribed as 'Thom'' (or specific form 
    seen). (Rule D.55)
*   'Robertus/um' abbr: MUST be transcribed as 'Rob't' (or specific form 
    seen). (Rule D.57)

*(Note: The following list provides standard forms. Your transcription MUST 
reflect the visual form on the image line, prioritizing the Critical 
Abbreviations above and Rule C.9.)*
11. **'&' (CRITICAL):** Transcribe Tironian 'et' as '&'.
12. **'et' / 'Et' (CRITICAL):** Transcribe spelled-out 'et' or 'Et' exactly as 
    written (matching case).
# ADDED: Explicit negative constraint for & / Et
**CRITICAL REMINDER:** **NEVER** transcribe a spelled-out 'Et' or 'et' on the 
image as '&'. **NEVER** transcribe a Tironian '&' on the image as 'Et' or 
'et'. Follow the image exactly.
13. **'&c':** Transcribe 'etcetera' abbreviation (e.g., '&c', '&c.') as '&c'.
14. **'q'd':** Transcribe 'quod' abbreviation (e.g., 'q' w/ macron or similar 
    mark) as 'q'd'.
15. **'quod':** Transcribe spelled-out 'quod' as 'quod'.
16. **'p'':** Transcribe 'per'/'pro' abbreviation (crossed-p) as 'p''.
17. **'per'/'pro':** Transcribe spelled-out 'per' or 'pro' as written.
18. **'com'':** Transcribe 'com'/'con' prefix abbreviation (macron/hook) as 
    'com''.
19. **'comitatus':** Transcribe spelled-out 'comitatus', 'comiti', etc., as 
    written.
20. **'d'ni'/'d'no':** Transcribe 'dominus/i/o' abbreviation as 'd'ni' or 
    'd'no'.
21. **'Dominus':** Transcribe spelled-out 'Dominus', 'domini', 'domino' as 
    written (respect case).
22. **'Ioh'es'/'Ioh'em':** Transcribe 'Johannes/em' abbreviation as 'Ioh'es' 
    or 'Ioh'em'.
23. **'Iohannes':** Transcribe spelled-out 'Iohannes', 'Iohannem' as written.
24. **'vic'':** Transcribe 'vicecomes/ti' abbreviation as 'vic''.
25. **'vicecomes':** Transcribe spelled-out 'vicecomes', 'vicecomiti' as 
    written.
26. **'p'd'':** Transcribe 'predictus' forms abbreviation (e.g., 'p'dcus', 
    'p'dco') as 'p'd'cus', 'p'd'co', 'p'd'ca', etc.
27. **'predictus':** Transcribe spelled-out 'predictus', 'predicto', etc., as 
    written.
28. **'Reg'':** Transcribe 'Regis/Rege/Regi' abbreviation as 'Reg''.
29. **'Regis'/'Rege'/'Regi':** Transcribe spelled-out 'Regis', 'Rege', 'Regi' 
    as written (capitalized).
30. **'Angl'':** Transcribe 'Anglie' abbreviation as 'Angl''.
31. **'Anglie':** Transcribe spelled-out 'Anglie' as written.
32. **'Westm'':** Transcribe 'Westmonasterium' abbreviation as 'Westm''.
33. **'Westmonasterium':** Transcribe spelled-out 'Westmonasterium' as 
    written.
34. **'attorn'':** Transcribe 'attornatus/um' abbreviation as 'attorn''.
35. **'attornatus':** Transcribe spelled-out 'attornatus', 'attornatum' as 
    written.
36. **'nup'':** Transcribe 'nuper' abbreviation as 'nup''.
37. **'nuper':** Transcribe spelled-out 'nuper' as written.
38. **'ap'd':** Transcribe 'apud' abbreviation as 'ap'd'.
39. **'apud':** Transcribe spelled-out 'apud' as written.
40. **'sup'':** Transcribe 'super' abbreviation as 'sup''.
41. **'super':** Transcribe spelled-out 'super' as written.
42. **'saluo'':** Transcribe 'saluo' with an abbreviation mark as 'saluo''.
43. **'saluo':** Transcribe spelled-out 'saluo' as 'saluo'.
44. **'pl'ito':** Transcribe 'placito' abbreviation as 'pl'ito'.
45. **'placito':** Transcribe spelled-out 'placito' as 'placito'.
46. **'transgr'':** Transcribe 'transgressio...' abbreviation as 'transgr'' 
    (or 't'nsgr'' if seen).
47. **'ib'm':** Transcribe 'ibidem' abbreviation as 'ib'm'.
48. **'ibidem':** Transcribe spelled-out 'ibidem' as 'ibidem'.
49. **'scil''/'sc'l't':** Transcribe 'scilicet' abbreviation as 'scil'' or 
    'sc'l't' (matching image).
50. **'scilicet':** Transcribe spelled-out 'scilicet' as 'scilicet'.
51. **'Will'm':** Transcribe 'Willelmum/us' abbreviation as 'Will'm' (or 
    specific form seen).
52. **'Willelmus':** Transcribe spelled-out 'Willelmus', 'Willelmum' as 
    written.
53. **'Ric'':** Transcribe 'Ricardus/um' abbreviation as 'Ric'' (or specific 
    form seen).
54. **'Ricardus':** Transcribe spelled-out 'Ricardus', 'Ricardum' as written.
55. **'Thom'':** Transcribe 'Thomas/am/e' abbreviation as 'Thom'' (or specific 
    form seen).
56. **'Thomas':** Transcribe spelled-out 'Thomas', 'Thomam', 'Thome' as 
    written.
57. **'Rob't':** Transcribe 'Robertus/um' abbreviation as 'Rob't' (or specific 
    form seen).
58. **'Robertus':** Transcribe spelled-out 'Robertus', 'Robertum' as written.
*(Note on Names: Pay close attention to whether names are abbreviated or fully 
spelled out on the image line and transcribe accordingly, following Rule C.9.)*

---
**E. REFERENCE EXAMPLES (Transcribe ONLY if seen on image)**
*(This list shows common abbreviated phrases. Transcribe the exact form 
visible.)*
*   absq' hoc q'd, accion'm ... h'ere non debet, ad cognosc', ad dampnum / Ad 
    g've dampnu', ad faciend' & recipiend' q'd cur' ... cons', ad largum 
    dimittatur, ad sectam, ad valenc', armig', assumpsit p' se ip'o, q'd 
    attach' eu' / q'd att'o, bene et veru' est q'd, cal'pn' fu't, q'd cap'et 
    eu' / capiatur / capiat, capiend' inde explec' ad valenc', ciuis et ..., 
    clausum ... fregit, cl'icus, cons' est q'd, Et cont' pacem d'ni Regis, 
    cuius dat' est die et Anno sup'dcis, de Com' in Com', de die in diem, de 
    pl'ito q'd redd', de pl'ito detenc'o'is catallor', de pl'ito quare ... de 
    novo fecit, defend' ius ... quando etc. / defend' vim et iniur' quando 
    etc., p' defalt', deteriorat' sunt, diem p' ess' suos, die impetrac'onis 
    b'ris originalis, domu' ... fregit, eat inde sine die, Et alia enormia 
    etc., Et h'et etc., Et hoc paratus est verificare, Et hoc petit q'd 
    inquiratur p' p'riam, Et ip'e non ven', Et p'd'cus ... similiter, Et q'd 
    tale sit ius suum offert etc., Et saluo &c., Et totum etc., Et unde &c., 
    Et vic' modo mand', exigat' eu' in Hustengo, q'd exigi fac' / exigifac', 
    execut' test'i, expediens & necesse est, fenu' inde p'venient' ... cepit 
    et asportavit, fil' et hered', gentilman' / Gentilman, gratis ... 
    warr'izat, h'as d'ni Regis de p'donat'o utlagat'e, h'eat de t'ra ... ad 
    valenc', husbondman, Idem dies dat' est, imp'p'm, in d'nico suo ut de 
    feodo et iure, in m'ia / in misericordia, in Octab's s'ti Hillar' / a die 
    pasche in xv dies / etc., in pp'ia p'sona sua, Io' / I'o, Ita q'd h'eret 
    corpus eius hic / Ita q'd h'eat corpus eius hic, iuxta formam statuti, 
    latitat vagat' et discurrit, p' legem t're, licet sepius requisit', 
    manucep'unt, m'cator / m'cer, nich'il c'piat p' br'e suu', nich'il h'et / 
    nichil habet, non est invent', non est p's / non s't p'sec', non obstante 
    p'allegato, nondum reddidit, nup' de, nup' maiore', op' se iiij die, 
    ostens' si quid ... quare ... non debeat, pandoxator, p' aliqua p'allegata 
    ab accio' sua ... p'cludi non deb', p' attornatu' suu', p' br'e d'ni Reg' 
    de recto, p'c' fuit vic' / prec' est vic' / p'cept' fuit vic' / p'cept' 
    est vic', pet' iud'm / petit iudiciu', pet' iud'm de br'i, pet' licenc' 
    inde int'loquendi / vlt'ius pet' licenc', pet' recogn' fieri, pet' v'sus, 
    pon' se in magnam assi'am d'ni Reg', postq'm sum' etc., p' int'esse suo, 
    p'munt' fuit essond', p'muniant' p'usq'm, pbos & leg'les ho'ies, Et 
    p'ferunt hic in cur' sc'ptum p'dcm, p'ut patet t'mio... Ro, p'ut p' b're 
    et narracom' sup'pon', quare vi et armis, que fuit ux', queritur de, 
    quiete de, quiet' & exon'at' a cur' dimittatur, quousq' &c., p' quos etc 
    Et qui nec etc Ad recogn' etc Quia tam etc., recogn' de t'ris & catallis 
    suis ad opus d'ni Regis levari, recup'et seisinam suam, scil't, scire 
    fac', p' quoddam sc'ptum suu' obligator', s'viens, set in contemptum cur' 
    recessit et defalt' fec', set sit in m'ia p'ro fa'l clam', si &c., sicut 
    plur' / sicut plur'ies / sicut p'us, solempnit' exact', soluend', sub 
    pena, q'd sum' eu' / q'd sum' eos, tenend' sibi et her' suis, tenentem p' 
    warr' suam, t'pore pacis t'pore d'ni Reg' nunc, unde p'duc' sectam etc., 
    ut ius et hereditatem suam, utlagat' / utlaget', utrum ip'e maius ius 
    h'eat, uterq' eor' sum' est p', vic' non mis' br'e, voc' inde ad warr', 
    yoman'

---
**F. LETTERFORMS & SPELLING**
59. **CRITICAL LETTERFORMS (U/V & I/J):** **ALWAYS** use only 'u' or 'U' 
    (never 'v' or 'V') and only 'i' or 'I' (never 'j' or 'J'). This applies to 
    all words, including 'uersus', 'uilla', 'ualenciam', 'iudicium', 
    'iuratus', etc.
60. **Long S:** Transcribe long 's' as standard 's'.
61. **Differentiation:** Carefully distinguish minims (n/u, m/in/ni/iu/ui), 
    c/t, and f/s based on the image. **# ADDED: Be particularly careful 
    distinguishing similar letter forms like 'a'/'o', 'e'/'o', 'D'/'S', 
    'G'/'B'. Always verify against the image.**
62. **Standard Letters:** Transcribe other letters to standard modern 
    equivalents based on image form.
63. **'uersus'/'sicut':** Transcribe 'uersus'/'u'sus' and 'sicut'/'sic'' based 
    on image form (abbreviated or full).

---
**G. CAPITALIZATION (Based on Image)**
64. **Strict Following (CRITICAL):** Follow manuscript capitalization 
    **EXACTLY** for **ALL** words.
    *   *Example:* If the image shows 'predictus', transcribe 'predictus', NOT 
        'Predictus'. If the image shows 'Comes', transcribe 'Comes', NOT 
        'comes'. If the image shows 'die', transcribe 'die', NOT 'Die'. If 
        image shows 'Anno', transcribe 'Anno', NOT 'anno'. Match the image 
        case **exactly** for **all** words. # CHANGED: Strengthened rule and 
        added more examples.
65. **'Rex'/'Dominus'/'Regis':** Capitalize 'Rex', 'Dominus', 'Regis', 'Rege', 
    'Regi' if spelled out on image.
66. **Nouns/Titles:** Capitalize proper nouns (people, places like 'London'', 
    'Pasche'), titles/occupations ('Gentilman', 'yoman', 'clericus') *only if 
    capitalized on the image*. Retain abbreviation marks if present.

---
**H. NUMERALS & SYMBOLS**
67. **Roman Numerals:** Transcribe Roman numerals exactly as they appear 
    (e.g., 'xij', 'xv', 'iiij'); do NOT convert to Arabic.
68. **Paragraph Mark:** Transcribe the paragraph mark symbol as '¶' if 
    present.

---
**I. WORKFLOW & VERIFICATION**
69. **Image Focus:** Examine the image, identifying the **red numbers** 
    marking the start of each requested baseline.
70. **Structure Analysis:** Note paragraphs and line counts using the **red 
    image numbers** and Named Entity List for context. Use the Named Entity 
    List as a reference for potential names/places, but **# ADDED: always 
    prioritize the spelling, capitalization, and form seen on the image** for 
    your transcription.

71. **PRELIMINARY STEP (Internal Grounding - DO NOT OUTPUT THIS): LAST WORD 
    FOCUS**
    *   For each **requested line number** (e.g., 10, 11, ... corresponding to 
        original document lines N, N+1, ...):
        *   Identify the *last word* written in the HTR transcription for that 
            line (use HTR only as a guide here).
        *   Carefully examine the *end* of the corresponding baseline on the 
            **Document Image Chunk**, identified by its **red number**.
        *   Determine the *correct* last word for that baseline based *solely* 
            on the **image**. Pay close attention to its spelling and any 
            abbreviation marks.
        *   **Crucially:** Mentally note this image-correct last word. This 
            word *must* remain as the final word on its corresponding line 
            (identified by its red number and matching JSON key) in the final 
            'abbreviated_latin_lines' output. **This step is critical to 
            prevent words from incorrectly wrapping to the next line.**

72. **Transcribe 'abbreviated_latin_lines' (FINAL OUTPUT):**
    *   Go line-by-line according to the **red baseline numbers** requested 
        for the chunk ('n').
    *   Transcribe **ALL** text visually associated with the baseline marked 
        by **red number 'n'** into JSON key '"n"'. **Do not omit any words or 
        symbols.** # CHANGED: Added emphasis on ALL text.
    *   Preserve exact image abbreviations, spelling, and capitalization found 
        on that specific numbered line.
    *   Use apostrophe (') only for visible marks/omissions per Rule 10.
    *   Apply all relevant rules (especially B-H and I.71-73) based *only* on 
        image evidence for that specific numbered line.
    *   **Ensure the last word identified in the Preliminary Step (Rule 71) is 
        correctly placed at the end of its line and NO words are moved across 
        lines.**
    *   **Crucially, ensure you ONLY transcribe lines corresponding to the 
        requested red numbers. Ignore any other text visible.**
    *   **Self-Correction Check (Per Line):** For each line 'n' you 
        transcribe, quickly verify: Does the text in JSON key '"n"' start with 
        the first word on the image baseline 'n'? Does it end with the last 
        word on the image baseline 'n'? **Does it include ALL words/symbols 
        visible on that baseline?** If not, re-read the image for that line 
        and correct it before moving to the next line. # CHANGED: Added check 
        for completeness.

73. **Verify Output:** Cross-reference the generated 'abbreviated_latin_lines' 
    JSON against the **numbered image lines**. Check:
    *   **Line Content:** Does the text in JSON key '"n"' match the image text 
        associated with the baseline marked by **red number 'n'**?
    *   **Completeness (CRITICAL):** Does the transcribed line contain *all* 
        the words/symbols visible on the image baseline? Are *any* words 
        missing?
    *   **Word Order:** Is the word order identical to the image for that 
        numbered line?
    *   **Line Breaks (CRITICAL):** Does the *first* word of JSON key '"n"' 
        match the *first* word on the image baseline marked 'n'? Does the 
        *last* word of JSON key '"n"' match the *last* word on the image 
        baseline marked 'n'? **Are there absolutely NO words moved between 
        lines?**
    *   **Correct Lines Transcribed:** Does the JSON contain keys *only* for 
        the requested line numbers, and is the text derived *only* from those 
        specific numbered lines, ignoring any other visible lines?
    *   **Abbreviation Forms (CRITICAL):** Are abbreviations transcribed 
        *exactly* as seen on the specific numbered line, following Rule C.9 
        (NO EXPANSION) and the Critical Abbreviations list in Section D?
    *   **Rule Adherence:** Are all other rules (capitalization, I/J/U/V, 
        etc.) followed based *only* on the image evidence for that numbered 
        line?

---
**FINAL REMINDER:** Ensure your final response is ONLY the JSON object 
containing the 'abbreviated_latin_lines' dictionary. Adhere strictly to 
all rules, especially:
1.  **Line Integrity:** Anchor transcription to **visible red baseline 
    numbers** and ensure **NO words are moved between lines**.
2.  **Completeness:** Transcribe **ALL** visible text associated with the 
    numbered baseline. **DO NOT OMIT** any words or symbols.
3.  **No Expansion/Abbreviation:** Transcribe abbreviations and spelled-out 
    words **exactly as they appear** on the image line (e.g., 'Et' stays 'Et', 
    '&' stays '&', 'Will'm' stays 'Will'm', 'Willelmum' stays 'Willelmum').
4.  **Exact '&' vs 'Et':** Follow the image precisely. **NEVER** convert 
    spelled-out 'Et'/'et' to '&', or '&' to 'Et'/'et'.
5.  **Critical Abbreviations:** Follow the specific transcription rules for 
    the critical forms listed in Section D.
6.  **Exact Capitalization:** Match manuscript capitalization **exactly** for 
    **ALL** words.
7.  **Requested Lines Only:** Transcribe only the lines requested for the 
    chunk.
\end{Verbatim}

\section{LLM Prompts for Case Study Analysis}

\subsection{LLM Expansion Prompt}
\label{app:expansion}
\begin{verbatim}
You are an expert in Latin paleography, specializing in the transcription of 
15th-century English legal records. Your task is to produce a full, scholarly 
transcription from a highly abbreviated text. Task: Take the provided abbreviated 
Latin text and expand it into a complete and readable version. You must strictly 
adhere to the editorial conventions listed below. Editorial Conventions and 
Transcription Rules: Distinguish between the vowel i and the consonant j. Use j 
for the consonantal sound (e.g., judicium, Johannes, injuriam). Distinguish 
between the vowel u and the consonant v. Use v for the consonantal sound (e.g., 
venit, verificare, vi) and u for the vowel sound (e.g., suum, unde). Always 
expand the ampersand (&) and the Tironian et to the full word et. Expand the 
abbreviation &c to etc. (with a period). Expand all other scribal abbreviations 
to their full form. Consistently use medieval Latin spelling conventions, using e 
instead of the classical ae diphthong. Expand abbreviations for "aforesaid" as 
predictus and its variants (never praedictus). 
Expand common legal formulas using standard phrasing, such as "ab actione sua 
predicta precludi non debent" and ensure correct word separation like "in nullo".
\end{verbatim}

\subsection{LLM Named-Entity Correction (NEC) Prompt}
\label{app:nec}
\begin{verbatim}
ROLE:
You are an expert historical researcher specializing in medieval and early modern 
Latin paleography, onomastics, and historical linguistics. Your primary strength 
is in identifying subtle HTR errors that less experienced analysts might miss. 
You are particularly skilled at questioning internally consistent but potentially 
erroneous HTR outputs, especially for proper nouns involving ambiguous capital 
letters.

CONTEXT:
* Source: The text is a direct output from HTR software applied to a handwritten 
  document. It is expected to contain errors.
* Document Type: Plea roll
* Time Period: 15th century
* Geographical Location: [e.g., Hertfordshire, England]
* Paleographical Watch-List: Based on the secretary hand script of this period, 
  you should pay special attention to potential confusions between capital letters 
  (such as B/G, S/F, W/M) and common minuscule confusions (such as c/t, u/n/m, 
  and long-s/f).
* Language: Latin, with potential vernacular (Middle English) words.

TASK:
Please perform the following two-part analysis.

Part 1: Detailed Entity Analysis
First, identify all named entities in the text. For this analysis, a "named 
entity" includes:
* Persons: Forenames and surnames.
* Places: Towns, counties, manors, specific fields.
* Organizations: (If any, e.g., guilds, abbeys).
* Significant Proper Nouns: Unique named objects (like a ship or a specific 
  cask of wine), and formal temporal markers (like feast days).
For each named entity, provide the following structured analysis:
* Entity Name: State the name clearly as a header.
* Textual Variants: List all different spellings or forms of this entity 
  exactly as they appear in the text.
* Plausible Alternatives & Probabilities: Acknowledge that the HTR text may be 
  flawed. Suggest plausible alternatives, considering paleographical ambiguities. 
  Assign a probability to each possibility (including the text's reading), 
  ensuring the total sums to 100%.
* Most Likely Reading & Justification: Determine the single most likely reading 
  and justify your choice, weighing onomastic plausibility against paleographical 
  likelihood.

Part 2: Rerendered Text
After completing your analysis, provide a final, clean version of the text.
* Substitute: Rerender the entire text, substituting each entity with your 
  determined "Most Likely Reading".
* Indicate Strong Alternatives: If any proposed alternative has a probability of 
  5% or higher, include it in square brackets '[like this]' immediately following 
  the most likely reading upon its first appearance.
\end{verbatim}

\end{document}